\documentclass[aps,pra,twocolumn,superscriptaddress,showpacs,nofootinbib]{revtex4-1}
\usepackage{amsmath}
\usepackage{amssymb}
\usepackage{graphicx}
\usepackage{verbatim}
\usepackage{xcolor}
\usepackage{epstopdf}
\usepackage{bbm}
\usepackage{url}
\usepackage{tikz}
\newcommand*\circled[1]{\tikz[baseline=(char.base)]{
		\node[shape=circle,draw,inner sep=0.5pt] (char) {#1};}}

\begin{document}

\title{Deep learning and high harmonic generation}

\author{M. Lytova}
\email[]{mlytova@gmail.ca}
\affiliation{National Research Council of Canada, 100 Sussex Drive, Ottawa ON K1A 0R6, Canada}
\affiliation{Department of Physics, University of Ottawa, Ottawa, Canada, K1N 6N5}

\author{M. Spanner}
\email[]{michael.spanner@nrc.ca}
\affiliation{National Research Council of Canada, 100 Sussex Drive, Ottawa ON K1A 0R6, Canada}
\affiliation{Department of Physics, University of Ottawa, Ottawa, Canada, K1N 6N5}
 
\author{I. Tamblyn}
\email[]{isaac.tamblyn@nrc.ca}
\affiliation{National Research Council of Canada, 100 Sussex Drive, Ottawa ON K1A 0R6, Canada}
\affiliation{Department of Physics, University of Ottawa, Ottawa, Canada, K1N 6N5}
\affiliation{Vector Institute for Artificial Intelligence, Toronto, Ontario, Canada}

\date{\today}

\begin{abstract} 
Using machine learning, we explore the utility of various deep neural networks (NN) when applied to high harmonic generation (HHG) scenarios.  First, we train the NNs to predict the time-dependent dipole and spectra of HHG emission from reduced-dimensionality models of di- and triatomic systems based of on sets of randomly generated parameters (laser pulse intensity, internuclear distance, and molecular orientation). These networks, once trained, are useful tools to rapidly generate the HHG spectra of our systems.  Similarly, we have trained the NNs to solve the inverse problem - to determine the molecular parameters based on HHG spectra or dipole acceleration data.  These types of networks could then be used as spectroscopic tools to invert HHG spectra in order to recover the underlying physical parameters of a system.  Next, we demonstrate that transfer learning can be applied to our networks to expand the range of applicability of the networks with only a small number of new test cases added to our training sets.  Finally, we demonstrate NNs that can be used to classify molecules by type: di- or triatomic, symmetric or asymmetric, wherein we can even rely on fairly simple fully connected neural networks.  With outlooks toward training with experimental data, these NN topologies offer a novel set of spectroscopic tools that could be incorporated into HHG experiments.
\end{abstract}

\maketitle

\section{Introduction}

High harmonic generation (HHG) is a multiphoton excited-state process which occurs in molecules and solids. Initiated by the interaction of a strong and short laser pulse with materials in the gas, liquid, or solid phase, HHG was first observed over 40 years ago \cite{Burnett1977}. In the years that followed, it received many experimental confirmations and theoretical developments, see e.g. \cite{McPherson, Ferray, Chang97, Chang99, Krausz}.  Although the key features of the process can be captured within a semi-classical model \cite{Corkum93}, a more accurate theoretical treatment includes solving the time-dependent Schr\"odinger equation (TDSE), see e.g.  \cite{Lewenstein, Zuo}. Such advanced numerical models typically applied in this area are usually computationally expensive \cite{Kolesik2004, LCB2007,Illustrations}.

Concurrently, the past decade has seen rapid improvements of the capabilities of artificial intelligence (AI) and machine learning (ML) methods. Application of such techniques within the physical sciences have been very fruitful, resulting in the acceleration and improved scaling of computational methods \cite{C8SC04578J}, the discovery of new materials \cite{Raccuglia}, and the design of synthetic pathways for molecules \cite{Gottipati}. To date there have been few examples of ML applied to the problem of HHG, with existing examples being the use of neural networks (NN) to predict high-harmonic flux under various experimental conditions in gas \cite{NN_HHG1} and plasmas \cite{NN_HHG2}, and to recover the pulse shape of attosecond bursts of radiation \cite{PhaseRetrieval}.  Deep learning in particular may offer a route to the theoretical treatment of materials and interfaces where traditional \emph{ab initio} based methods are either intractable or very expensive to calculate. 

In this paper, we explore a variety of ML scenarios as applied to computations of HHG phenomena.  We first consider the prediction of HHG emission given a set of molecular and laser parameters by training a network with a set of precomputed HHG emissions obtained by repeated numerical solution of the TDSE.  Once trained, these neural networks offer a rapid method to predict the HHG spectra for a range of input parameters that circumvents the need for further numerical solution of the TDSE. Second, we consider the inverse problem where we extract the underlying molecular parameters given the HHG emission.  These types of NNs could eventually be applied to experimental HHG data in order to recover the underlying physical parameters of the systems being studied.  Next, we demonstrate the use of transfer learning to enlarge the range of applicability of a trained neural network with a minimal number of new training examples.  From a theoretical perspective, transfer learning will be crucial in expanding networks trained on an easily-computable dataset to include example of cases that are much harder to compute, for example a network trained on two-dimensional simulations could be expanded by including a small number of example from three-dimensional cases.  Alternatively, a network trained on experimental data for one molecule could be expanded by including a small number of examples from a new one.   Finally, we demonstrate a NN that can identify the type of molecule (i.e.  diatomic, triatomic, symmetric, asymmetric, etc) that resulted in a given HHG emission, similar to the classic machine learning problem of classifying pictures of cats and dogs, which offers a novel spectroscopic tool that ultimately could be applied to, for example, HHG probing of chemical transformations.

This article is organized as follows. Section II provides a short introduction to relevant AI/ML methods and notation.  We discuss several different neural network architectures, including their data and training requirements, highlighting their key capabilities as they relate to time-series data such as those produced in HHG.  In Section III we discuss the numerical computation of the datasets used for training our neural networks.  In Section ~\ref{ML_exp}, we apply the techniques from Section II to simulated prototypical HHG signals, thereby demonstrating the potential for deep learning to be used as an enabling technology within the field. Links to detailed Jupyter notebooks of all models and training procedures are available at the end of this section.  Section~\ref{Outlooks} concludes the paper and offers direction and perspectives for future application of deep learning to the field of HHG.

\section{Deep learning}\label{DL}

Deep learning is a sub-field of machine learning which operates directly on ``raw'' observations without human-controlled feature engineering or selection.  The ability of deep neural networks to self-learn the importance of different hierarchical features within a training set makes it ideal for application to physical phenomena, where there is often a competition between processes which occur across a range of time- and length-scales. The similarity between the learned multi-scale coarsening within a deep neural network and real-space renormalization group has been noted \cite{Beny}. 

There are many books which cover the various objectives and models of ML \cite{Goodfellow-et-al-2016}; this Section is not intended to cover them in full. However, we need to clarify our notation and explain what is meant by each type of model we use in Section~\ref{ML_exp}. Since our ML experiments are mainly applied to time series, here we first use periodic waves (e.g.  $\sin(t)$) for pedagogical illustrations. 

\subsection{Surrogate model}\label{surr}

A surrogate model is a data-driven model based on observations from a physical system, numerical simulator, or analytical equations which seeks to reproduce behaviors of the reference system at reduced cost. Simply put, a surrogate model should, once trained, replicate $(\textrm{input}, \textrm{output})$ pairs of the reference system. Surrogate models are often used during iterative design processes \cite{Koziel2011, Keane}, developing dynamical control schemes \cite{Biannic}, and route planning \cite{Queipo}. Neural networks have been shown to be flexible and efficient function approximators and hence are now often used as surrogate models.

\begin{figure}[t]
	\includegraphics[width=0.95\columnwidth]{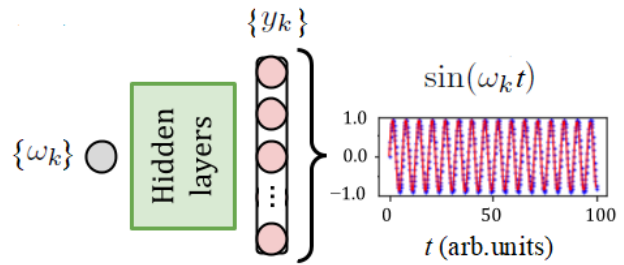}
	\caption{Surrogate model for predicting $\sin(\omega_kt)$ by given value of $\omega_k$. The target function is shown with blue dots, the prediction with red line. \label{surr_fig}}
\end{figure}

As an example, suppose we want to solve the harmonic oscillator problem with a NN. We have as observations a set of randomly generated frequencies $\omega_k\in[0.5,\: 1]\; (k=1,2,\dots,N_{train})$, and the corresponding set of solutions $y_k(t_n)=\sin(\omega_kt_n)$ on the grid $t_n\in[0,\:T] \;(n=1,2,\dots,N_{g})$. From these sets, we can train a NN consisting of one-node input layer ``$\omega$'', an output layer ``$y$'' with number of nodes equal to $N_{g}$, and some number of hidden layers (of various sizes and architectures) in-between to compute $y_k(t_n)$ by given $\omega_k$ \textit{without} resorting to the built-in function $\sin()$. In other words, we build a \textit{surrogate} for the solution, which predicts value of $y$ for any \textit{test} input value $\omega\in[0.5,\: 1]$ at any point $t_n$, see Fig.~\ref{surr_fig}.  

In the case of a fortunate choice of the hidden layer architecture, the accuracy of our calculations increases with the number of passes (\textit{epochs}) of the whole dataset through the network. After each $m$-th forward propagation we measure the averaged (over all $k$ and $n$) difference between the values $\{y_k(t_n)\}^m$ predicted by the NN and the \textit{target} values $y_k(t_n)$ using some \textit{loss function}. Depending on the value of this error, the weights of the NN are corrected at the $m$-th stage of backpropagation of the error, thus representing an example of \textit{supervised} learning.  Obviously, in practice, it is interesting to construct an accurate surrogate model (using as small a training sample as possible) to reproduce a multiparameter function that is difficult to compute using analytical or numerical methods.

\subsection{Regression model}

\begin{figure}[t]
	\includegraphics[width=0.95\columnwidth]{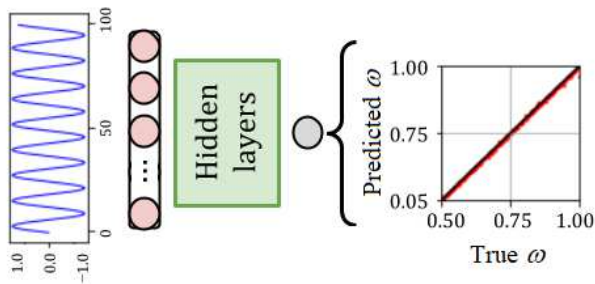}
	\caption{Regression model for predicting frequency of sine wave based on $y_k(t_n)$ dataset. On the right: plot of the ``True vs. Predicted'' values of $\omega$ for the entire test sample. \label{reg_fig}}
\end{figure}

Now suppose that with the same training sets as in Section~\ref{surr} one wants to teach another NN to determine the $\omega_k$ parameter that gives the best fit for $y_k(t_n)$ in the $(y, t)$ plane. In statistics, this type of model is called regression \cite{Casella}. In the case of a sine wave we deal with nonlinear regression. There are a huge number of regression models based on both the prior assignment of the dependence $y(t_n)$ of a certain form (like polynomial regression, e.g. \cite{scikit-learn}) and on the feature learning methods such as kernel ridge regression \cite{Faber,Lopes,Brockherde_2017} and random forests \cite{Faber,Ward2016}. However, here we use a term ``regression model'' specifically for the NN that is trained to find the parameters (in the sine wave example, only ``$\omega$'') associated with the input function $y(t_n)$, see Fig.~\ref{reg_fig}. The regression model is also an example of supervised learning where we use a loss function with respect to the target value of $\omega_k$. In practice, such a model may be useful to determine parameters related to experimental data.  

\subsection{Transfer learning}\label{transfer}

Suppose we have a trained surrogate model for calculating the function $y_k(t_n) = \sin(\omega_kt_n)$ based only on a given value of $\omega_k$. Using a \emph{transfer learning} technique we can adapt our model to calculate some closely related function, say, $y_k(t_n) = 2\sin(\omega_kt_n)$, and we do not need to create a new surrogate model from scratch. Instead, we freeze all the weights of the hidden layers in the original model, except for those immediately following the input. Training such a \textit{pre-trained} model (in fact, only weights for one layer) requires a much smaller dataset and training time \cite{tensorflow2015, chollet2015keras}.

\subsection{Classifying model}

To categorize some objects into $N$ classes, we first need to assign each of them a label, for example, an integer from 0 to $N-1$. Suppose we are given a training set of labeled instances, whereas the classes (labels) of test instances are unknown. The task is to build an algorithm capable of classifying the latter. 

In the classification of objects it is assumed that they have some features. A set of these features related to one object is called its feature vector. Thus, to classify an object means to indicate the label to which this object belongs with the highest level of confidence, depending on its vector of features.  The problem is quite common, so there are many ML approaches to solving it, e.g.  \cite{Voulgaris, Wang2010}. What we call the classifying model here is a deep learning approach, without explicit specification of the features.  ``Deep'' in application to NN means we are using multiple hidden layers which are capable of learning hierarchical features. The labeled time series are fed to the input of the neural network, which determines for itself the features characteristic of each class.

Similar to surrogate and regression models, the classifying model is also an example of supervised learning.  As the neural network is trained, the loss function between the predicted and true results determines the necessary adjustment of the weights for the current stage. The result of using the classifying model for a test instance is an array of length $N$, whose elements contain the probabilities that this instance belonging to each particular class \cite{tensorflow2015}.

\section{Calculating datasets with TDSE}\label{tdse_S}

We consider the generation of high harmonics from model one-electron molecules driven by an ultrafast strong linearly polarized laser field ${\bf E}(t) = {\bf E}_0\sin(\omega_0 t)$ with frequency $\omega_0 = 0.057$ atomic units (a.u.), which corresponds to a wavelength of 800 nm. We neglect interactions between molecules (i.e. we assume that the medium is dilute), and solve the TDSE:
\begin{equation}\label{tdse}
	i\partial_t\psi({\bf r},t)=\hat{H}({\bf r},t)\psi({\bf r},t).
\end{equation} 
For simplicity we restrict \eqref{tdse} to two spatial dimensions: ${\bf r} = (x,y)$. We consider di- and triatomic molecular models defined collectively by the Hamiltonian (in atomic units)
\begin{equation}\label{Ham}
	\hat{H}(x,y,t) = -\dfrac{1}{2}\left(\partial_x^2+\partial_y^2\right) + V_C(x,y,R)+xE_0\sin(\omega_0t),
\end{equation}
where the electric field is directed along the $x$-axis, and the binding potential of nuclei is
\begin{eqnarray}\label{Coulomb}
	\nonumber &&V_C(x,y,R)=-\dfrac{q_1}{\sqrt{x'^2+(y'-\frac{R}{2})^2+\varepsilon^2}}-\\
	&&\dfrac{q_2}{\sqrt{x'^2+(y'+\frac{R}{2})^2+\varepsilon^2}}-
	\dfrac{q_3}{\sqrt{(x'-\frac{\sqrt{3}R}{2})^2+y'^2+\varepsilon^2}},\quad
\end{eqnarray}
The parameter $R$ allows us to vary the bond lengths.  We restrict our models to have a single unit of effective total charge ($q_1+q_2+q_3 = 1$) such that we mimic ionization of neutral molecules.  In the case of diatomic molecules we set $q_3=0$. Note that with the Hamiltonian Eq.(\ref{Ham}) all triatomic molecules have symmetric bondlengths (the expression assumes that $R$ is the same for both bonds of the triatomic molecules), but we can introduce asymmetry by choosing unequal values of the charges $q_i$.  On the right side of \eqref{Coulomb}, the new coordinates $(x',y')$ are obtained from $(x,y)$ using the appropriate rotation matrix, so that $q_1$ and $q_2$ both lie on the $y'$-axis. The phenomenological parameter $ \varepsilon=\varepsilon(R)$ is used to adjust the ground state energy of the system.  In our simulations we use
\begin{equation}
	\varepsilon(R)=\left\{\begin{array}{l}
	-0.21R+0.78,\; \text{for}\; R\in[1.5, 3),\\
	-0.09R+0.42,\; \text{for}\; R\in[3, 4]
	\end{array}
	\right.
\end{equation}
which, for convenience, ensures that the ground state energy and hence ionization potential does not vary too strongly as we vary the molecular geometries.  Cartoons of the resulting di- and triatomics generated by our Hamiltonian can be see below in Fig.\ref{Exp3}.  We solve the TDSE \eqref{tdse} numerically using the split-operator spectral method \cite{Feit82}.  The simulations were run for 7.25 cycles of the 800 nm laser starting at a zero of the field oscilations with the laser intensity held constant. The HHG emission was calculated via the dipole acceleration ${\bf \ddot d}(t)$ using Ehrenfest's theorem
\cite{Burnett1992}, 
\begin{eqnarray}\label{dxdy}
\nonumber&&(a_x(t), a_y(t))^T\equiv(\ddot{d}_x(t), \ddot{d}_y(t))^T =\\
\nonumber 
 &&\int\!|\psi(x,y,t)|^2(\partial_x V_C(x,y,R)+E(t),\partial_yV_C(x,y,R))^Tdxdy.\\
\end{eqnarray}
Further in the text, we refer to $\ddot{d}(t)$ assuming it is $x$-component from \eqref{dxdy}. We call the Fourier transform of that component the spectral intensity of the high harmonics: $S(\omega)=|\hat{ \ddot{d}}_x(\omega)|^2$.

\begin{figure*} 
	\includegraphics[width=0.99\textwidth]{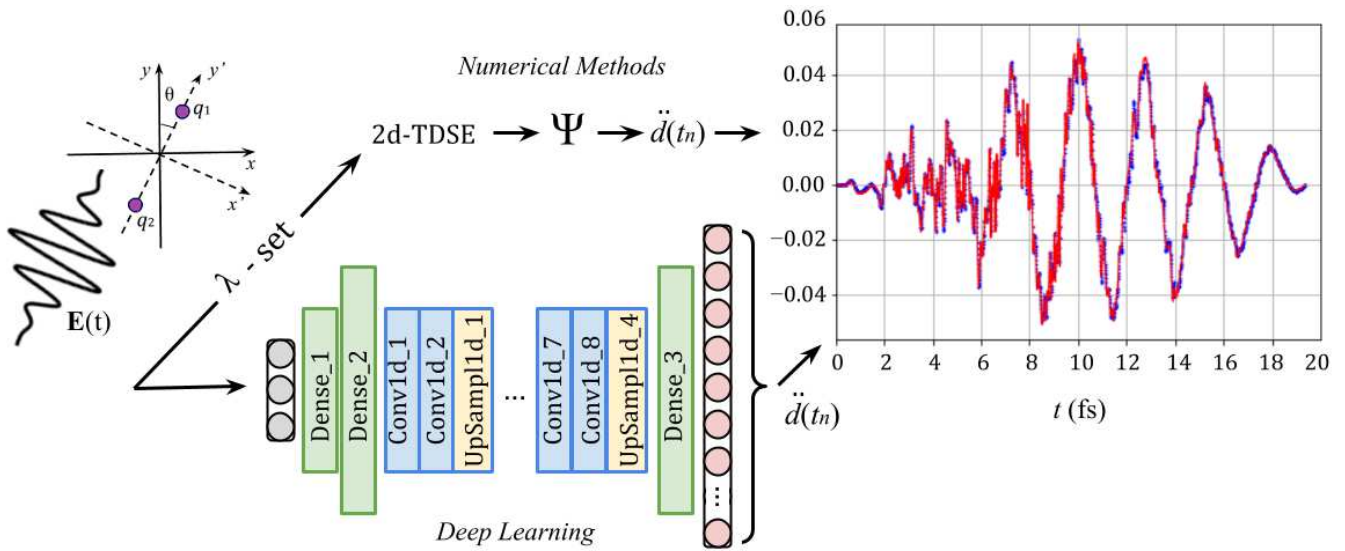}
	\caption{\label{Exp1} We feed the $\lambda$-set of parameters of the symmetric diatomic molecule: $(q_1 = q_2 = 1/2)$ to the NN input and compare the outputted dipole acceleration $\ddot{d}(t_n)$, $n=1,2,\dots,4096$ (curves shown in red) with $\ddot{d}(t_n)$ computed through the numerical solutions of the 2D-TDSE (dots shown in blue). The hidden part of the NN consist of several fully-connected (dense) layers and four blocks of 1D-Convolutional layers accompanied by 1D-UpSampling layers. The activation function for layers is the hyperbolic tangent. Parameters for the presented example: $\theta = 50.13^\circ $, $R = 3.67 $ a.u., $I = 1.64\cdot10^{14}$ W$/$cm$^2$; the dipole acceleration units are Hartree a.u., $N_{train}=30000$. The loss function is the mean squared error (MSE). }
\end{figure*} 

\section{Computational experiments with neural networks}\label{ML_exp}

We use a collection of the numerical solutions of 2D-TDSE as datasets to train the NNs for various purposes. Let us rewrite \eqref{tdse} with the Hamiltonian
\eqref{Ham} as
\begin{equation}
i\partial_t\psi = \hat{H}(\lambda)\psi,
\end{equation}
where we omitted the notation for the variables $(x,y,t)$ involved in the partial derivatives, although emphasized the presence of a set of parameters $\lambda = \{\theta, R, I\}$. In this set, $\theta$ is an angle between the axis of molecule and the electric field, and $I=E_0^2$ is the laser peak intensity. The $q_i$ dependence of the Hamiltonian appears below as distinct cases of molecules. To train the NNs, we first generate datasets of size $N_{train}$ in the time or frequency domain on the grid $(n = 1,2,\dots, N_{g})$ : $\{\ddot{d}_k(t_n)\}$, $\{S_k(\omega_n)\}$ $(k=1,2\dots N_{train})$ for one, two or all three parameters of $\{\lambda_k\}$-set randomly distributed in their intervals: (i)~$\theta\in[0^\circ,\; 90^\circ]$; (ii)~$R\in[1.5,\; 4]$ a.u.; (iii)~$I\in[1,\;4]\times10^{14}$ W$/$cm$^2$.  In our experiments, we use the TensorFlow \cite{tensorflow2015} and Keras \cite{chollet2015keras} libraries.

\subsection{Surrogate of the dipole acceleration}\label{gen_d}

First, we train NNs to calculate the dipole acceleration $\{\ddot{d}_k(t_n)\}$ from the set of independent random parameters $\{\lambda_k\}$. The principal idea of this ML experiment is shown in Fig.~\ref{Exp1}.

\subsubsection{Symmetric diatomic molecule}\label{sym_di}

To start, we fix the internuclear distance at $R = 2$~a.u., and the laser intensity at $I = 1.5 \cdot10^{14}$~W$/$ cm$^2$, leaving only the angle $\theta$ to change randomly from 0 to $90^\circ$.  For this and many of the following trainings in the paper, we use the Adam optimizer with learning rate $lr =5\cdot10^{-4}$, unless otherwise declared.  Here the training set size is large $N_{train}=10,000$. We address the large data requirements subsequently to transfer learning. Specifically, for the $\lambda$-set with two fixed parameters, the entire learning process takes 200 epochs, with the final value of the mean-squared error (MSE) approaching $10^{-8}$ (the estimated error $\leq 0.01\%$). With such a small final MSE, true and predicted results cannot be discerned by eye.

Next, we train the same architecture NN to calculate $\{\ddot{d}_k(t_n)\}$ from the set $\{\lambda_k\}=\{\theta_k, R_k, I_k\}$, all of which change randomly and independently in the intervals given above. In this case we observe much more diverse shapes of the dipole acceleration curves. In particular, since the ratio of maximal and minimal peak intensities varies quite a bit ($I_{max}/I_{min}=4$), the amplitudes of $\ddot{d}_k(t)$ spread over 2 orders of magnitude. As a result, the training requires a longer processing time (compared to the case when only angle $\theta$ changes randomly, while other parameters are frozen) and we need to use a bigger training set $N_{train}=30,000$. Over several of thousand epochs of trainings with increasing batch size \cite{Smith18} and repeated training cycles, MSE still does not fall below the value $10^{- 5}$. Despite the fact that MSE is several orders of magnitude higher than in the case of only one changing parameter, from the example HHG plotted in Fig.~\ref{Exp1} it can be seen that even in this situation, the trained model is able to capture important features of the shape of the dipole acceleration.

\begin{figure}[b]
	\includegraphics[width=0.99\columnwidth]{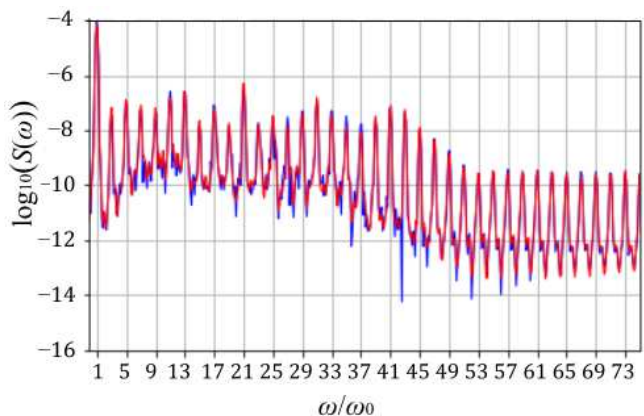}
	\caption{An example of the spectral intensity prediction from the set $\theta = 13.65^\circ $, $R = 2.36 $ a.u., $I = 2.49\cdot10^{14}$ W$/$cm$^2$.  The spectrum numerically computed through the TDSE is shown in blue, while the NN prediction is given in red. Activation function - softplus. For the training set of size 30000, the final MSE $=6\cdot 10^{- 4}$.\label{thRI_spec}}
\end{figure} 

Considering the prediction of the dipole acceleration from a set of parameters using NN as not only accurate but also computationally efficient method, we compared it with N-D linear interpolation methods \cite{scipy}. Our calculation showed that although in the 2D-case (time $t_n$ and, for example, the angle $\theta$ variables) the results can be obtained even faster using the interpolation methods, however, with the addition of new parameters and hence increasing dimensionality of interpolation, the latter method becomes less effective, especially taking into consideration that for the sake of accuracy, we want to keep the number of nodes in time $N_{g}\geq 4096$.

The general answer to the question of how accurately it makes sense to predict the functions $\ddot{d}(t_n)$ is not obvious. For example, if we train a NN to predict the corresponding spectral intensity $S(\omega_n)$, we can reasonably limit our demands to accurate predicting of the harmonics maxima on the plateau and near cut-off region, neglecting the fact that the MSE itself is not very small, see Fig.\ref{thRI_spec}.

\subsubsection{Transfer learning for asymmetric diatomic and symmetric triatomic molecules} 

In order to reduce the data requirements, so an approach is practical for real systems, we apply a ML technique known as transfer learning \cite{Goodfellow-et-al-2016} to train the models for other types of molecules.  The schematic diagram of the experiment is the same as shown in Fig.~\ref{Exp1}, however, instead of learning from scratch, we used a pretrained model for a symmetric diatomic molecule in which the weights were frozen for all layers except the last two fully connected layers (see Section~\ref{transfer}).  Using transfer learning, we re-optimize this NN model using a small number of training examples generated for asymmetric diatomics and symmetric triatomics.

\begin{figure}[b]
	\includegraphics[width=0.99\columnwidth]{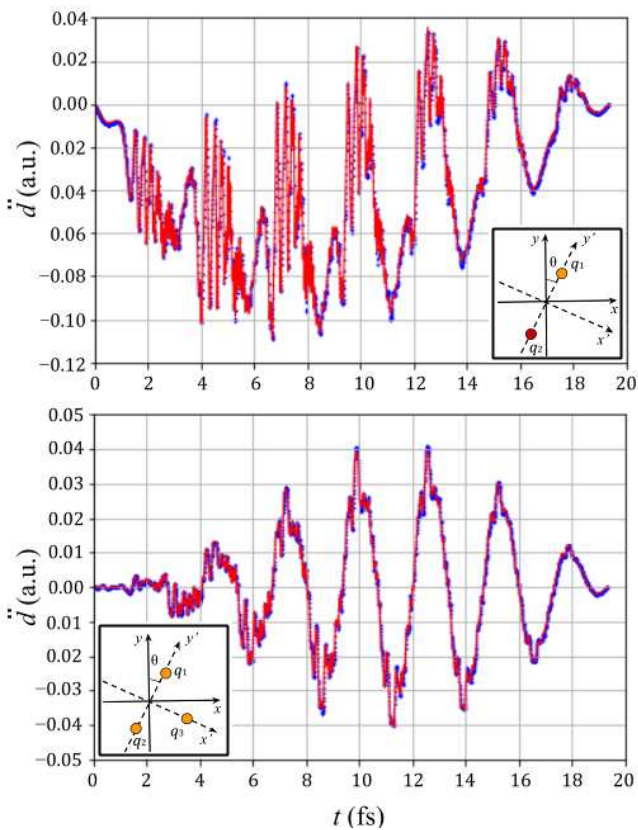}
	\caption{Two examples of predicting the acceleration of the molecular electron using the transfer learning technique. Top: asymmetric diatomic molecule $(q_1 = 1/3,\; q_2 = 2/3)$, parameters: $\theta = 73.77^\circ$, $R = 3.2$ a.u., $I = 2.5\cdot10^{14}$ W$/$cm$^2$. Bottom:  symmetric triatomic molecule $(q_1 = q_2 = q_3=1/3)$, parameters: $\theta = 65.61^\circ$, $R = 2.04$ a.u., $I = 1.67\cdot10^{14}$ W$/$cm$^2$.  The test functions (solutions obtained with TDSE) are shown in blue dots, the model predictions are given in red. Activation function - $\tanh$. The training sample size is 5000 for both examples. \label{transf1}}
\end{figure}

In the case of $\ddot{d}(t_n)$ prediction, the method allowed us to use for training $\times$6 smaller datasets: 5,000 vs 30,000 used for model described in section~\ref{sym_di}. Fig.~\ref{transf1} demonstrates two examples obtained within that approach. By thus reducing the size of the training sample (and hence the processing time), we nevertheless achieve the same MSE $\approx 10^{-5}$ as in the case of the original model for a symmetric diatomic molecule. Moreover, we can even reduce the size of the training set for transfer learning down to 2000 if we train the NN to predict the spectral intensity $S(\omega_n)$ instead of predicting time-dependent dipole acceleration. In the latter case, we need $\times$4 fewer points in frequency, and our requirements to the accuracy of predicting for the shape of the spectra could be less stringent than in the case of $\ddot{d}(t_n)$, as we discussed at the end of Section~\ref{sym_di}.  

\begin{figure*}
	\includegraphics[width=0.92\textwidth]{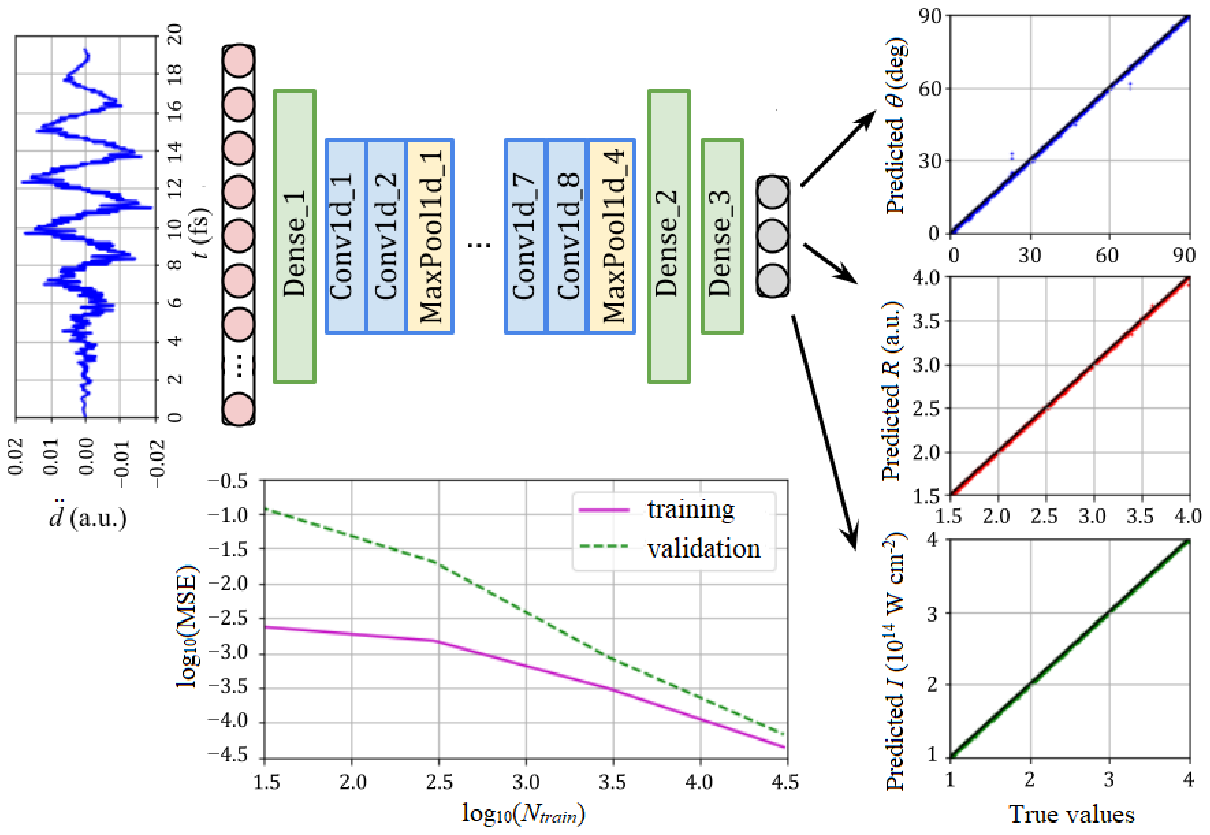}
	\caption{\label{Exp2} Training of the regression NN model to estimate the parameters of TDSE solutions. We feed $\ddot{d}(t_n)$, $n=1,2,\dots,4096$ (blue curve to the left) to the NN input, and compare the outputted parameters with the corresponding ``true'' parameters of the $\lambda$-set used to solve the TDSE for the respective $\ddot{d}(t_n)$. The hidden part of the NN consist of several fully-connected (dense) layers and four blocks of 1D-Convolutional layers accompanied by 1D-MaxPooling layers. The activation function for layers is the hyperbolic tangent. The loss function is the MSE.  Bottom center shows the model's learning curve as the training sample size increases: $N_{train} = \{30, 300, 3000, 30000\}$}
\end{figure*} 

\subsection{Estimating of molecular and laser parameters}\label{L_estim}

In this section, we demonstrate that by feeding a dataset of $\ddot{d}(t_n)$ vectors computed via TDSE to the NN input, we can train this regression network to extract the corresponding set of molecular and laser parameters $\{\theta, R, I \}$. Fig.~\ref{Exp2} presents the scheme of this training experiment. To estimate the accuracy of prediction we rely on the loss function final values (MSE $\leq 10^{-4}$ after 2000 epochs and for $N_{train}=30000$) and visually compare the ``true'' parameters (which were used to compute the input dataset $\{\ddot{d}_k(t_n)\}$ via the TDSE) and the values predicted by the NN. Of the three inserts on the right in Fig.~\ref{Exp2} we can conclude that the model gives a fairly accurate prediction.  Bottom-centered learning curves (for training and validation datasets) demonstrate how the MSE decreases with increasing of the training set size, so that the mean error $\sqrt{\text{MSE}}$ finally becomes $\leq 1\%$. Similar regression experiments for the $\lambda$-parameters can also be made based on the spectra of higher harmonics.

Note that, as in the previous section, the inclusion of several convolutional layers is an important feature of our modeling, as it allows NNs to be trained on datasets with tens of thousands of training examples instead of hundreds of thousands, which would be necessary in a deep but fully dense architecture to achieve the same level of the MSE by the end of the training process. It is known that convolutional networks are distinguished by a very high ability to recognize patterns not only in images and handwritten characters \cite{Lecun}, but also in time series data.

Finally, we apply transfer learning to extend this parameter-extracting NN to include asymmetric diatomic molecules. The accuracy of such a retrained model turned out to be lower than for the original model predictions: $\sqrt{\text{MSE}}=3.8\%$. We observe that the prediction quality is worse, despite the fact that we have frozen the internal (convolutional) weight of the model and retrained all the fully connected layers. It seems that these convolutional layers are especially important in coding the angle $\theta$ parameter, so their excluding from the training process can result in noticeable errors.

\subsection{Distinguishing diatomic and triatomic molecules by HHG emission}

\begin{figure*}
	\includegraphics[width=0.98\textwidth]{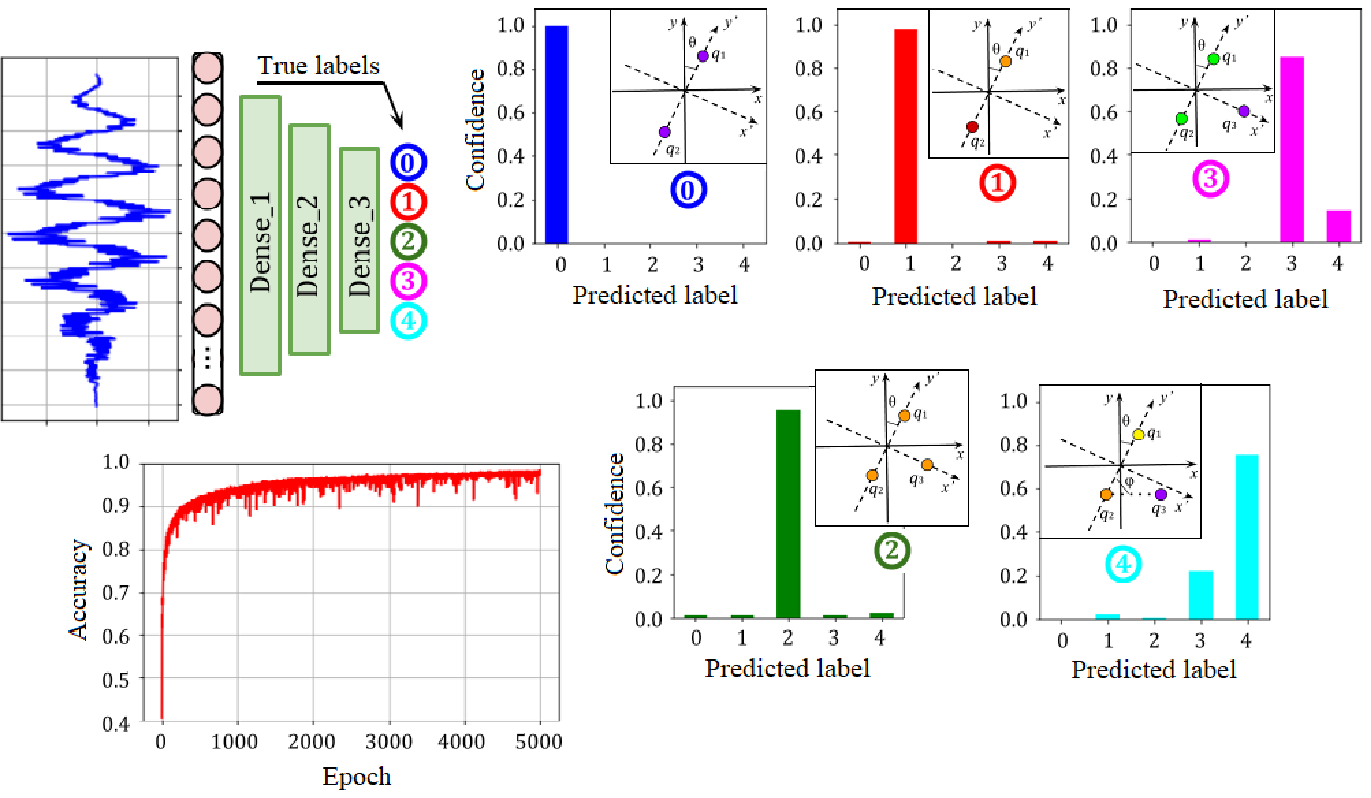}
	\caption{\label{Exp3} Top left corner: a neural network model to classify the molecular electron accelerations $\ddot{d}(t)$ (computed through 2D-TDSE) by five species of the respective diatomic and triatomic molecules. The training accuracy curve for $N_{train}=5000$ is shown at the bottom left. On the right side, the predicted distributions calculated on the test sample of 500 are presented. Inserts in the diagrams show molecular geometry for the corresponding input/true labels, see text and Tab.~\ref{geometry}}
\end{figure*} 

In the previous sections we demonstrated that the convolutional NNs can be trained to successfully reproduce the shapes of the dipole acceleration curves and spectra from the set of parameters as well as extract these parameters from the functions $\ddot{d}(t)$ or $S(\omega)$.  In addition to the values of specific parameters, we also assume that the molecules have a specific geometry and various effective charges of positive centers. In Tab.~\ref{geometry} we gathered parameters related to five types of molecule under consideration. This time we denote the internuclear distance between $q_1$ and $q_2$ as $R_1\in[1.5, 4]$ a.u. For triatomic molecules we introduce $R_2$ as a distance between vertices $q_2$ and $q_3$. We are considering three kinds of triatomic molecules, however for one of them, marked as \circled{2}, $R_2$ is chosen to be equal to $R_1$. Also note that the $\varphi$ angle between $R_1$ and $R_2$ changes randomly only in the case marked as \circled{4}. As before, $\theta\in[0^\circ, 90^\circ]$ and $I\in[1,\;4]\times10^{14}$~W$/$cm$^2$.

\begin{table}[t]
	\caption{Geometrical parameters for models of di- and triatomic molecules.\label{geometry} }
	\begin{ruledtabular}
		\begin{tabular}{cccccc}
			Label                   & $q_1$    & $q_2$               & $q_3$ & $R_2$ (a.u.)    & $\varphi$ \\ \hline
			\circled{0} & 1/2 & 1/2 & 0       & $-$           & $-$           \\
			\circled{1} & 1/3  &2/3  & 0   & $-$       & $-$         \\
			\circled{2} & 1/3  & 1/3 & 1/3     & $R_1$       & $60^\circ$          \\
			\circled{3} & 1/4  & 1/4 & 1/2     & $[3,5]$       & $\arccos{(R_1/2R_2)}$          \\
			\circled{4} & 1/6  & 1/3 & 1/2     & $[3,5]$       & $[45^\circ, 75^\circ]$          \\
		\end{tabular}
	\end{ruledtabular}
\end{table}

In this part, we investigate the question of whether NN can classify molecules by their types based on HHG emission.  Fig.~\ref{Exp3} represents the NN we train.  The input to this network is 4096 points in time $d(t_n)$, which is followed by three hidden fully connected layers of 128, 64 and 16 nodes with rectified linear unit (ReLU) activation.  The output layer consists of 5 nodes, so for each input, the result is an array of 5 numbers in the $ [0,1] $ interval that represent the confidence for each possible outcome. We train the model using the Adam optimizer with standard learning rate $= 10^{-3}$ and the probabilistic loss function Sparse Categorical Crossentropy: 
\begin{equation}
	CCE = -\dfrac{1}{N}\sum_{i=1}^N\sum_{j=1}^M \mathbbm{1}_{ y_i\in C_j}\log[p(y_i\in C_j)],
\end{equation} 
where $\{y_i\}$ $(i=1,2,\dots N)$ denotes a dataset, $C_j$ are number of classes under consideration $(j=1,2,\dots M)$, $\mathbbm{1}$ is the indicator function, and $p$ is the model probability for the $y_i$  to belong to the class $C_j$.

In addition to the NN architecture, Fig.~\ref{Exp3} also shows the model predictions on the testing dataset. It can be seen that symmetric diatomic \circled{0}, asymmetric diatomic \circled{1}, and symmetric triatomic \circled{2} species differ with a high degree of confidence. If we wanted to distinguish only these three types, then it would take only 200 epochs, moreover, it would be enough to use two hidden layers instead of three.  However, the presence of the species \circled{3} and \circled{4} complicates the task and requires up to 5000 epochs to achieve confidence of at least $75-80\%$. A possible reason is that for random $R_2$ and $\varphi$, the potentials in the last two cases become very close, especially since $q_1$ and $q_2$ differ by very little ($\pm \frac{1}{12}$) between these cases. All this leads to close and hardly distinguishable dipole accelerations $\ddot{d}(t)$. Further training does not help here, we can increase the prediction accuracy only by increasing the size of the training sample (here it is 1000 for each species).  Finally, note that such a model can also be applied to classify molecules by their HHG spectra (which is readily available in experiments) instead of the dipole acceleration.

\subsubsection*{Data availability}

All codes used in Sections~\ref{DL} and \ref{ML_exp} are available here
\url{http://clean.energyscience.ca/codes} and the training datasets are here
\url{http://clean.energyscience.ca/datasets}

\section{Outlooks}\label{Outlook}

Machine learning methods, which are very successful in a wide range of scientific and technological applications, have to date not been broadly applied to the field of high harmonic generation. This work aims to fill this gap and proposes new methods for studying this phenomenon using the machinery of deep learning.  

We have demonstrated a number of neural network architectures that can be useful in studying HHG. Once trained, the NNs can be used to predict the HHG emission given an set of system and laser parameters.  This will alleviate the need for repeated numerical solution in situations were rapid HHG computations are required, for example when propagating a Maxwell equation interacting with a HHG medium where the HHG emission must be computed at many positions and intensities concurrently.  The NNs can also solve the inverse problem where the systems parameters can be rapidly estimated from an HHG spectrum, which offers a novel tool for HHG spectroscopy.  We have demonstrated that transfer learning can be successfully applied to these NNs in order to expand the range of applicability by retraining the NNs with a relatively small number of new test cases.  Such techniques will be crucial to grow the NNs by showing them new training examples generated from either additional computational results or new experimental data.  In analogy with the classic dog vs cat task of machine learning, we have successfully generated NNs that can classify the type of molecule that leads to a given HHG spectrum.  These classifying NNs again represent new spectroscopic tools that could be used to quickly identify the presence or absence of particular species when using HHG as a probe of chemical transformations.

\section{Acknowledgments}
M.S. and I.T. acknowledge financial support from the Natural Science and Engineering Research Council (NSERC) of Canada through their Discovery Grants program.

\bibliography{dl_hhg_v2}

\end{document}